\documentclass{article}

\usepackage{hyperref, amsmath, graphicx, capt-of}

\newcommand*{\SectRef}[1]{Section \ref{#1}}
\newcommand*{\FigRef}[1]{Fig.~\ref{#1}}
\newcommand*{\Action}{\mathcal{S}}

\begin{document}
\title{\sc Cosmological modelling with Regge calculus}
\author{Rex G Liu and Ruth M Williams\\
{\small DAMTP, CMS, Wilberforce Rd  Cambridge, CB3 0WA, UK} \\
{\small {\em Email} : R.G.Liu.01@cantab.net}
}

\date{}

\maketitle

\begin{abstract} 
The late universe's matter distribution obeys the Copernican principle at only the coarsest of scales.  The relative importance of such inhomogeneity is still not well understood. Because of the Einstein field equations' non-linear nature, some argue a non-perturbative approach is necessary to correctly model inhomogeneities and may even obviate any need for dark energy.  We shall discuss an approach based on Regge calculus, a discrete approximation to general relativity: we shall discuss the Collins--Williams formulation of Regge calculus and its application to two toy universes. The first is a universe for which the continuum solution is well-established, the $\Lambda$-FLRW universe. The second is an inhomogeneous universe, the `lattice universe' wherein matter consists solely of a lattice of point masses with pure vacuum in between, a distribution more similar to that of the actual universe compared to FLRW universes. We shall discuss both regular lattices and one where one mass gets perturbed.
\end{abstract}

\section{Introduction}

Friedmann--Lema\^itre--Robertson--Walker (FLRW) models have been extremely successful in explaining many cosmological observations, including, most notably, the Hubble expansion, the cosmic microwave background (CMB), and baryon acoustic oscillations.  Indeed, the underlying Copernican assumption appears well-supported by precision measurements showing the CMB to be isotropic to within one part in $10^5$ \cite{CMB-isotropy}.  Yet the late, matter-dominated universe obeys the Copernican principle at only the coarsest of scales: most matter is clustered into large-scale structures with large voids in between, and the physical effects of such `lumpiness' are still not fully understood.

There has been intense interest recently over the possible importance of inhomogeneities to observational cosmology.  For instance, it has been proposed that the universe's observed acceleration is actually an apparent effect, a result of fitting a homogeneous model to data from an inhomogeneous source, and because of its non-linear structure, the inhomogeneities can only be modelled correctly by non-perturbative methods \cite[and references therein]{dark-energy}.

Regge calculus \cite{Regge} offers one non-perturbative approach.  It is a discretisation of general relativity that can in principle approximate any space--time.  This work will focus on a form of Regge calculus first devised by Collins and Williams (CW) \cite{CW} and further developed by Brewin \cite{Brewin}.  We shall examine the formalism's potential to non-perturbatively approximate inhomogeneous cosmologies.  \SectRef{Regge} will present a cursory introduction to Regge calculus and the CW formalism.  To deepen our understanding of the formalism, \SectRef{Lambda} will apply it to a space--time for which the continuum solution is well-known, the closed vacuum $\Lambda$-FLRW space--time.  \SectRef{Lattice} will adopt the formalism to model a toy inhomogeneous universe, the closed `lattice universes' where matter consists of massive particles arranged into a regular lattice on space-like Cauchy surfaces.  Such matter content would be more representative of the actual universe's compared to that of FLRW.  \SectRef{Concl} will summarise and present a few directions in which this work can be extended.

\section{Regge calculus and the Collins--Williams formalism}
\label{Regge}

In general relativity, one can obtain the Einstein field equations by varying the Einstein--Hilbert action
\begin{equation}
\Action_\text{EH} = \displaystyle{\frac{1}{16\pi}\int R \, \sqrt{-g} \; d^4x}
\label{Einstein-Hilbert}
\end{equation}
with respect to the metric tensor $g_{\mu\nu}$, where $R$ is the Ricci scalar and $g = \det (g_{\mu\nu})$.  The action is evaluated over a continuous manifold.

\begin{figure}[htb]
\centering
\begin{minipage}{.47\textwidth}
  \centering
  \includegraphics{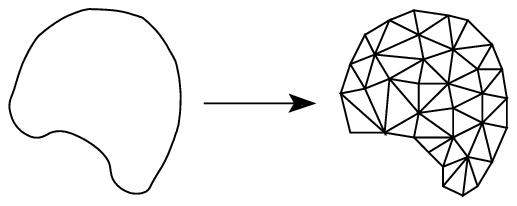}
  \captionof{figure}{Discretisation of a continuous manifold into a skeleton.}
  \label{skeleton}
\end{minipage}
\hspace{1mm}
\begin{minipage}{.47\textwidth}
  \centering
  \includegraphics{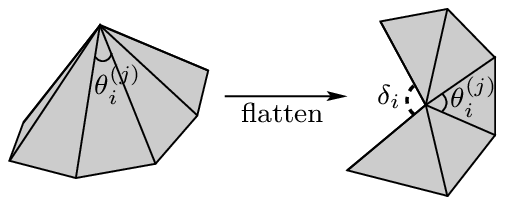}
  \captionof{figure}{Conical singularity at a 2-dimensional hinge (vertex). If the hinge is flattened, a gap of angle $\delta_i$ opens up between two blocks.}
  \label{hinge}
\end{minipage}
\end{figure}

The key idea of Regge calculus is to replace this continuous manifold with a discrete piece-wise linear one (\FigRef{skeleton}), known as a \emph{skeleton}.  A skeleton essentially consists of flat blocks glued together at shared faces.  As the blocks are flat, the interior metric is the Minkowski metric.  Curvature manifests itself as conical singularities centred on the sub-faces of co-dimension 2, known as \emph{hinges}; if one were to flatten a hinge (\FigRef{hinge}), a gap would open up between two of the blocks, and the size of this gap, the \emph{deficit angle}, gives a measure of the curvature.  The Regge analogue to the metric are the block edge-lengths.  Since curvature has support on the hinges only, the integration in \eqref{Einstein-Hilbert} reduces to a discrete summation over the hinges, and the Einstein--Hilbert action reduces to the \emph{Regge action}
\begin{equation}
\Action_\text{Regge} = \displaystyle{\frac{1}{8\pi}\sum_{i \,\in\, \left\{ \text{hinges}\right\}} \mkern-9mu A_i\, \delta_i},
\label{ReggeAction}
\end{equation}
where $A_i$ are the hinge areas and $\delta_i$ their corresponding deficit angles.  The \emph{Regge equations} are then obtained by varying $\Action_\text{Regge}$ with respect to the edge-lengths.

The CW formalism is essentially a Regge approximation of FLRW space--times using a skeleton that mimics FLRW symmetries.  FLRW space--times admit a foliation by Cauchy surfaces of constant curvature; all surfaces are identical to each other apart from an overall scale factor.  CW Cauchy surfaces approximate their FLRW counterparts by tessellating them with a single regular polytope.  Although the formalism can be generalised to other tessellations and background curvatures, we shall focus only on tessellations of closed universes using identical, equilateral tetrahedra; there are only three such possible tessellations, consisting of 5, 16, or 600 tetrahedra \cite{Coxeter}.  All edge-lengths in a surface are identical to each other, and all surfaces are identical to each other apart from an overall scaling.  To complete the 4-dimensional skeleton, the surfaces are glued together by a series of time-like \emph{struts} $\{m_i\}$ that connect vertices in one surface to their time-evolved images in the next.  Then the world-tubes of the tetrahedra between pairs of consecutive surfaces form the skeleton's 4-blocks.  The surfaces can be parametrised by a time parameter $t$.  After we have obtained the Regge equations though, we shall take the continuum time limit where the time separation between surfaces goes to zero, $dt \to 0$; this generates a series of differential equations for the Cauchy surface edge-lengths $l(t)$.  Thus, the CW formalism is actually a continuum time formulation of Regge calculus.

Since the original FLRW Cauchy surfaces are 3-spheres, our CW Cauchy surfaces are actually triangulations of 3-spheres and can therefore be embedded into 3-spheres in $\mathbf{E}^4$.  As Brewin noted \cite{Brewin}, the embedding radius $R(t)$ provides a natural analogue to the FLRW scale factor $a(t)$.  We explored this idea further in \cite{Lambda-FLRW} and found multiple ways to define the radius -- we could take, for instance, the radius to the vertices or to the tetrahedral centres -- but regardless of the choice, the radius would always be related to the tetrahedral edge-length by some constant scaling $Z$,
\begin{equation}
R(t) = Z \, l(t).
\label{3-sphere-rad}
\end{equation}

Brewin \cite{Brewin} noted that there are actually two ways to vary the Regge action.  The first is to impose the symmetry constraints on the skeleton first -- that is, require all tetrahedral edges in a surface to have equal length and all struts between the same pair of surfaces to have equal length.  When we vary one edge, the constraints require that all other edges sharing the same length be varied simultaneously.  This is called \emph{global variation}.  The alternative approach is to first vary each edge individually and afterwards impose the constraints.  This is called \emph{local variation} and is more analogous to how the Einstein--Hilbert action is varied in standard general relativity; in that case, the action gets varied with respect to an unconstrained metric first, and the Copernican symmetries are imposed afterwards on the resulting Einstein field equations.  To locally vary the CW skeleton though, we need to fully triangulate the skeleton; otherwise, the geometry of the varied skeleton would not be well-defined \cite{Brewin, Lambda-FLRW}.  Triangulation is done by introducing a set of diagonal edges $\{d_i\}$ between Cauchy surfaces.

If the local and global actions are equivalent, then the global Regge equation can be related to the local one via a chain rule \cite{Brewin}: if we vary globally with respect to some edge $q$, we can express this variation as
\begin{equation}
\frac{\partial \Action}{\partial q} = \sum_i \frac{\partial \Action}{\partial l_i^\ell} \frac{\partial l_i^\ell}{\partial q} + \sum_i \frac{\partial \Action}{\partial m_i^\ell} \frac{\partial m_i^\ell}{\partial q} + \sum_i \frac{\partial \Action}{\partial d_i^\ell} \frac{\partial d_i^\ell}{\partial q},
\label{chain-rule}
\end{equation}
where the superscript $\ell$ denotes that the relevant edges are being varied locally.  It is immediately evident that any solution of the local Regge equations, $0 = \frac{\partial \Action}{\partial l_i^\ell} = \frac{\partial \Action}{\partial m_i^\ell} = \frac{\partial \Action}{\partial d_i^\ell}$, would also be a solution of the global equation, $0 = \frac{\partial \Action}{\partial q}$, but the converse is not necessarily true.  In the models we shall consider, the local and global actions are indeed equivalent \cite{Lambda-FLRW, Regge-lattice}.

Finally, Brewin noted certain analogies between the CW and the ADM formalisms \cite{Brewin}.  The tetrahedral edge-lengths $\{l_i\}$ determined the Cauchy surface 3-geometry and were therefore analogous to the 3-metric $^{(3)}g_{ij}$.  The time-like struts were analogous to the ADM lapse functions.  The diagonals were analogous to the ADM shift functions.  Thus, we shall call the Regge equations obtained from the tetrahedral edges the evolution equation, from the struts the Hamiltonian constraints, and from the diagonals the momentum constraints.  There are, however, certain caveats to this analogy \cite[and references therein]{RGL}.

\section{CW models of closed $\Lambda$-FLRW space--times}
\label{Lambda}

When there is a non-zero cosmological constant $\Lambda$, both the Einstein--Hilbert action and the Regge action acquire a volume term
\begin{equation}
\displaystyle{\frac{1}{16\pi}\int \left( R - 2\, \Lambda \right) \sqrt{-g} \; d^4x} \;\; \to \;\; \displaystyle{\frac{1}{8\pi} \left[ \sum_{i \,\in\, \left\{ \text{hinges}\right\}} \mkern-9mu A_i\, \delta_i \; - \mkern-26mu \sum_{i \, \in \, \left\{ \text{4-blocks} \right\} } \mkern-21mu \Lambda\, V^{(4)}_i \right]},
\end{equation}
where $\{V^{(4)}_i\}$ are the 4-block volumes.  The Regge model was studied extensively in \cite{Lambda-FLRW}.  From both local and global variation, the continuum-time Hamiltonian constraint was found to be
\begin{equation}
l^2 = 6\, \frac{N_1}{N_3 \, \Lambda} \frac{(2\pi - n\theta)}{\tan\left(\frac{1}{2}\theta\right)},
\label{Lambda-constraint}
\end{equation}
where $N_1$ and $N_3$ are, respectively, the numbers of tetrahedral edges and tetrahedra in a Cauchy surface, $n$ is the number of tetrahedra meeting at a tetrahedral edge, and $\theta$ is the dihedral angle between any 4-blocks meeting at the time-like hinges located between Cauchy surfaces.  It was shown that this equation satisfies the initial value equation at the moment of time symmetry, which, for $\Lambda$-FLRW, is the moment of minimum expansion.  It was also shown that the constraint is a first integral of the global evolution equation, implying that it alone is sufficient to determine the model's evolution.  The constraint is also a first integral of the local evolution equation but only if we also satisfy the momentum constraints.  Unfortunately, the momentum constraints are actually unphysical because the diagonals break the Cauchy surface symmetries.  Hence, local variation in the CW formalism is unviable.

\begin{figure}[htb]
\centering
\includegraphics{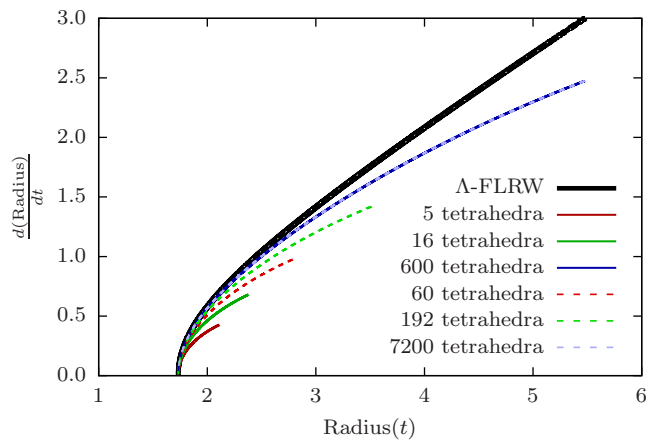}
\vspace{-3mm}
\caption{Evolution of Cauchy surface 3-sphere radii.  For continuum $\Lambda$-FLRW, this is simply the scale factor $a(t)$.  For the Regge models, we have chosen $Z$ in \eqref{3-sphere-rad} so that $R(t) = a(t)$ when $\dot{R}(t) = \dot{a}(t) = 0$.}
\label{Children-Lambda}
\end{figure}

The three Regge models' evolution have been plotted in \FigRef{Children-Lambda}; for comparison, the exact $\Lambda$-FLRW solution is also shown.  All Regge models correctly show an infinitely expanding universe, but all models also diverge slowly from the continuum solution as the universe expands; the divergence is slower though if the number of tetrahedra is higher.  This divergence arises from approximating an ever-expanding 3-sphere with just a fixed number of tetrahedra: as the 3-sphere expands, the approximation's resolution degrades, but the degradation is slower if there are more tetrahedra.


We can increase the number of tetrahedra by triangulating them into smaller tetrahedra, but the new tetrahedra will no longer be identical nor necessarily equilateral.  Brewin \cite{Brewin} has provided one subdivision algorithm that can be repeated indefinitely on each subsequent generation of child tetrahedra, but we shall focus only on the first-generation children.  Subdividing a parent Cauchy surface generates three different types of tetrahedral edges, tetrahedra, vertices, and struts.  We shall simplify the model by setting all strut-lengths to be equal.

This model was also investigated extensively in \cite{Lambda-FLRW}.  The Hamiltonian constraint, which we do not show, also satisfied the initial value equation at time symmetry.  But the constraint was a first integral of the evolution equation only if all tetrahedra were equilateral; otherwise, it was not a first integral in general, and we believe this to be a consequence of constraining all strut-lengths to be equal.  \FigRef{Children-Lambda} also shows the children models' evolution; our earlier remarks for the parents models apply to the children as well.  We also see that the rate of divergence depends only on the number of tetrahedra, independent of whether the skeleton is a parent or child.

\section{Lattice universes}
\label{Lattice}

If massive particles are present, the Einstein--Hilbert and Regge actions become
\begin{equation}
\displaystyle{\frac{1}{16\pi}\int R \, \sqrt{-g} \; d^4x \; - \mkern-25mu \sum_{i \, \in \, \left\{ \text{particles} \right\} } \mkern-25mu M_i \int ds_i} \;\; \to \;\; \displaystyle{\frac{1}{8\pi}\sum_{i \,\in\, \left\{ \text{hinges}\right\}} \mkern-20mu A_i\, \delta_i} \; - \mkern-35mu \sum_{\substack{i \, \in \, \left\{ \text{particles} \right\} \\ j \, \in \, \left\{ \text{4-simplices} \right\} }} \mkern-30mu M_i \, s_{ij},
\end{equation}
where $\{M_i\}$ are the particle masses, $s_i$ the length of particle $M_i$'s world line, and $s_{ij}$  the length of particle $M_i$'s world line through 4-block $j$.

In \cite{Regge-lattice}, the CW formalism was adapted to model lattice universes.  For each parent skeleton, there were different ways of arranging particles to form lattices, and it was found that the model's behaviour depended on the particle's position in the tetrahedron.  For regular lattices, we obtained the Hamiltonian constraint
\begin{equation}
l = 8 \pi M\, \frac{N_p}{N_1} \frac{\tan \frac{\theta}{2}}{\left[8 v^2 \tan^2 \frac{\theta}{2} - \frac{1}{2}\left(8v^2-1\right)\right]^\frac{1}{2}} \frac{1}{2\pi - n \theta},
\label{reg-constraint}
\end{equation}
where $N_p$ is the number of particles and $v$ the ratio between the particle's distance from the tetrahedron's centre and the tetrahedron's edge-length.  This constraint was also a first integral of the evolution equation.  We see from the constraint that evolution depends on the particle's position, parametrised by $v$, and it was shown in \cite{Regge-lattice} that the model would only be stable unconditionally if the particles were within a sphere that just touched the tetrahedral edge mid-points; this is believed to be an artefact of the Regge approximation rather than an actual feature of the lattice universes themselves.  We shall henceforth consider only lattices with the particles at the tetrahedral centres.

\begin{figure}[htb]
\centering
\includegraphics{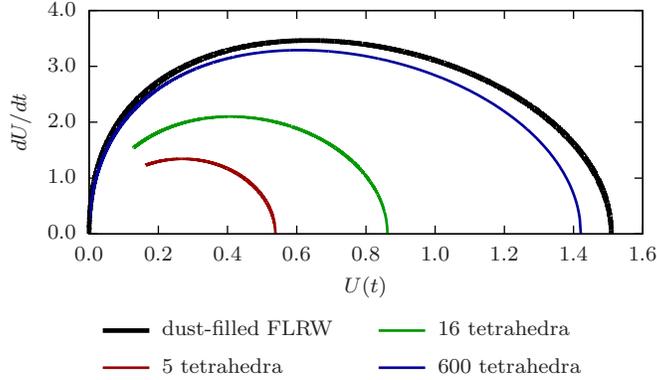}
\vspace{-7mm}
\caption{Evolution of the universe's volume $U(t)$ for lattice and dust-filled universes.  For the Regge models, $U(t)$ is the sum of the tetrahedra's volumes.}
\label{reg-lattice}
\end{figure}

\FigRef{reg-lattice} shows the evolution of regular lattice universes and of a dust-filled FLRW universe; the total mass is the same across all universes.  We see that the lattice universes are closed and stable.  They also become more similar to FLRW as the number of particles increases, and this is because their matter content is becoming more like homogeneous and isotropic dust.

We can perturb the lattice by changing one mass from $M$ to $M + \delta M$.  The skeletal geometry would then get perturbed, but depending on the skeleton, there could be anywhere from five to over a hundred independent tetrahedral edge-lengths.  Thus for simplicity, we shall focus only on the five-tetrahedra model, which involves just two independent lengths.

\begin{figure}[htb]
\centering
\includegraphics{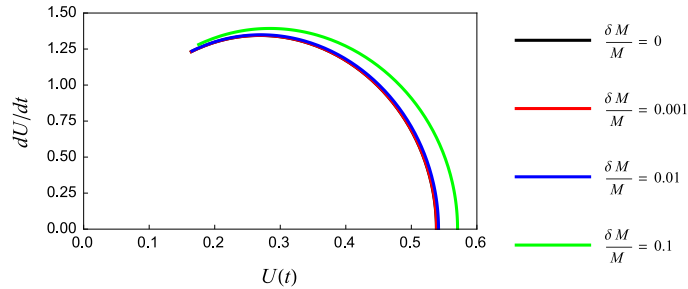}
\vspace{-3mm}
\caption{Evolution of the universe's volume $U(t)$ for various lattice perturbations.  The first three perturbations are so small that their graphs effectively overlap each other.}
\label{pert-lattice}
\end{figure}

The dynamics for this universe was also studied in \cite{Regge-lattice}.  It was now assumed the Hamiltonian constraint would be a first integral again of the evolution equation.  A global solution was this time obtained through local variation via the chain rule \eqref{chain-rule}.  The perturbed skeleton had two independent struts, so locally varying them gave two independent constraints in total; these were just enough equations to solve for the two independent tetrahedral edge-lengths.  Had we directly varied the action globally instead, we would have obtained just one constraint, which would not be enough.  In effect, the local approach allowed us to isolate one unique solution from the entire global solution space.  However, the two constraints were two coupled, non-linear differential equations for the two edge-lengths; therefore to solve the equations, we linearised them by taking their perturbative expansion in $\delta M/M$ up to first order.  The resulting equations were solved numerically, using as initial conditions the initial value equation at time symmetry, satisfied order-by-order in $\delta M/M$.  \FigRef{pert-lattice} shows the model's evolution for several perturbations.  In all cases, the evolution was stable; increasing the perturbation only increased the universe's maximum expansion.

\section{Conclusions}
\label{Concl}

Comparison with the exact $\Lambda$-FLRW solution indicates the CW formalism yields a reasonable approximation to cosmological space--times.  In particular, it reliably reproduces the universe's dynamics, and its accuracy increases with the number of tetrahedra.  It should be especially accurate for closed universes, as these do not expand indefinitely.

When applied to lattice universes, the formalism shows that both regular and perturbed universes have closed and stable evolution.  Yet we can take this work further in several ways.  Currently, our approximation of each lattice cell geometry is very coarse-grained, as there is only one independent edge-length characterising each cell.  But by subdividing the tetrahedra, we can introduce more edges and thereby increase the geometric detail.  We can also increase inhomogeneities by, for instance, having different cells with different masses or even leaving some empty altogether.  Finally, it would be especially interesting to investigate the models' optical properties and redshifts, as this may elucidate whether inhomogeneities can significantly affect cosmological observables.

\end{document}